\documentclass[aps,twocolumn,showpacs,superscriptaddress,citeautoscript,prb,floatfix]{revtex4-2}
\usepackage{graphicx}  
\usepackage{amsmath, amssymb, mathrsfs}
\usepackage{amsfonts}
\usepackage{tabularx}
\usepackage{cancel} 
\usepackage{bbold} 
\usepackage{wrapfig}
\usepackage{bm}
\usepackage[T1]{fontenc}
\usepackage{textcomp}
\usepackage{enumitem}
\usepackage{cancel}
\usepackage{bbold}
\usepackage{esvect}
\usepackage{commath}
\usepackage[most]{tcolorbox}
\usepackage{lipsum, babel}
\usepackage{comment}
\usepackage{lineno}
\usepackage{physics}
\usepackage{xcolor}
\usetikzlibrary{positioning,calc}
\tikzset{>=latex}
\usepackage[verbose,hypertexnames=false,bookmarksopenlevel=1,filecolor=blue,
linkcolor=blue,citecolor=blue,urlcolor=blue,pdfstartview=FitH,bookmarksopen,bookmarksnumbered,
colorlinks,plainpages=false,linktocpage]{hyperref}
\usepackage{cleveref}
\usepackage{adjustbox}

\usepackage[left=23mm,right=23mm, top = 20mm, bottom = 20mm, paper = a4paper]{geometry}
\begin{document}

\title{Tunable plasmon modes and topological transitions in single- and bilayer semi-Dirac materials}

\author{Debasmita Giri}
\affiliation{Institute for Theoretical Physics, University of Regensburg, D-93040
Regensburg, Germany }
\author{John Schliemann} 
\affiliation{Institute for Theoretical Physics, University of Regensburg, D-93040
Regensburg, Germany }
\author{Rafael Molina}
\affiliation{Instituto de Estructura de la Materia, CSIC,
Serrano 123, 28006 Madrid, Spain}
\author{Alexander Lopez}
 \email[To whom correspondence should be addressed. Electronic
 address: ]{alexander.lopez@physik.uni-regensburg.de}
\affiliation{Institute for Theoretical Physics, University of Regensburg, D-93040
Regensburg, Germany }
\affiliation{Escuela Superior Polit\'ecnica del Litoral, ESPOL, Departamento de F\'isica, Facultad de Ciencias Naturales y Matem\'aticas, Campus Gustavo Galindo
 Km. 30.5 Via Perimetral, P. O. Box 09-01-5863, Guayaquil, Ecuador}
 \affiliation{Departamento de F\'isica de Materiales, Universidad Complutense, 28040, Madrid, Spain}
 \begin{abstract}
 We investigate the plasmonic response of single- and bilayer semi-Dirac materials under the influence of a tunable parameter $\delta$ that governs topological transitions via Dirac cone generation/merging and incorporating band inversion terms. For single-layer systems, we demonstrate that the emergence of Dirac cones leads to an enhanced plasmon frequency range and that the plasmonic spectrum exhibits strong anisotropy, especially for finite $\delta$ and vanishing inversion terms.  In the bilayer configurations, we uncover a second plasmon mode whose relative phase, with respect to the first mode, can be actively controlled by rotating the upper layer which impacts the symmetry of the charge oscillations across the layers. This tunability enables switching between in- and out-of-phase plasmonic modes, offering a route toward phase-controlled collective excitations. Our results highlight the potential of semi-Dirac systems for topological plasmonics and interferometric applications in next-generation optoelectronic devices.
\end{abstract}
\maketitle
\section{Introduction}
Two-dimensional materials with a semi-Dirac (SD) Hamiltonian possess massless linear dispersion along one main axis and massive-like quadratic dispersion along the  perpendicular axis in momentum space, which makes them interesting candidates for developing devices with direction-dependent optical and transport properties. These materials were theoretically predicted to possess a $B^{2/3}$ magnetic field dependence of the Landau level energy spectrum \cite{sd1}, which interpolates between the linear relation for conventional two-dimensional massive fermions and the $B^{1/2}$ dependence observed in graphene for massless fermions. Further theoretical work \cite{pardo,pardoPRB,montambaux,MontambauxPRB2011,BanerjeePRB2012,MontambauxPRL2012,TarruelNature2012,MontambauxPRL2013,RudenkoPRB2014} has shown that type I semi-Dirac materials possess vanishing Chern numbers, whereas a Dirac cone merging condition can produce non-trivial Berry phases, via control parameters that can be assessed via potassium doping in black phosphorus\cite{MontambauxPRL2013,KimScience2015}.
The role of the Dirac merging condition in the diffusion of Dirac fermions is explored in reference\cite{AdroguerPRB2016} where the authors show that both the anisotropy of the Fermi surface and the Dirac nature of the eigenstates combine to give rise to anisotropic transport times, manifested through an unusual matrix self-energy. 
These SD materials can also support tilted cones
\cite{Zhang2017} coexisting in a striped of boron sheets or can be realized in silicene oxide
\cite{Zhong2017}. Moreover, the role of excitonic pairing has been addressed in an insulating transition in two dimensional SD semi metals\cite{WangPRB2017}, whereas valley selective Landau-Zener oscillations in SD p-n junctions have been reported in reference \cite{SahaPRB2017}.
The semi-Dirac features were experimentally observed in the material $NdSb$ \cite{NeupaneJPCM2016}, in black phosphorus with protection due to space inversion symmetry\cite{KimScience2017}, 
and, more recently, along the nodal lines of the topological three-dimensional metal $ZrSiS$\cite{Shao2024}, where the authors used  magneto- optical spectroscopy\cite{MurakamiPRB2022}.  Moreover, topologically distinct features have been predicted to arise between so called type I and type II semi-Dirac materials, which belong to different symmetry classes according to the classification of static topological materials \cite{MarimotoPRB2013}. The type II SD materials were proposed and shown to have a non-vanishing Chern number \cite{HuangPRB2015}.  Furthermore, type III Dirac cones that combine flat and linear dispersions were predicted in reference\cite{MilicevicPRX2019} for which Dirac cones emerge from the touching of a flat and a parabolic band when a synthetic photonic strain is introduced into the lattice.

Various asymmetric light transmission \cite{BorPRB2018} and 
signatures of merging Dirac points in optics and transport \cite{CarbottePRB2019-1,CarbottePRB2019}, as well as direction-dependent giant
 optical conductivity in two-dimensional semi-Dirac materials have also been reported recently\cite{Basky2019}. Berry curvature effects and 
 Hall viscosities in bulk anisotropic Dirac semimetal were reported in reference \cite{FranciscoPRB2019} and SD nanoribbon \cite{MurakamiPRB2020} where it has been 
shown that 
transport and anisotropic localization can emerge in polariton honeycomb lattices \cite{RealPRL2020}. 
The optical effects of the SD system have recently been explored; see \cite{NarayanPRB2015,SahaPRB2016,MohantaPRB2021,LiNJP2022,PeetersPRB2022} and references therein.   
Within this realm, the plasmon response in type I semi-Dirac material undergoing the topological transition associated with the Dirac cone merging is explored in reference \cite{ChakrabortyPRB2016}, where the authors show that the presence of the van Hove singularity in the electron spectrum leads to the existence of the gapped damped plasmon mode at zero chemical potential in the semimetal phase.
On the other hand, recent theoretical works explore the role of parity breaking mass terms with momentum dependence in the so-called Bernevig-Hughes-Zhang (BHZ) model, which can lead to  topologically non-trivial phases with finite Chern number\cite{Qi1PRB2006,Qi2PRB2006,Qi3PRB2006,BHZScience2006,MHZ} whose original proposal shows that the electronic state changes from  normal to inverted type with this band inversion being a topological quantum phase transition between a conventional insulating phase and a phase exhibiting the QSH effect with a single pair of helical edge states. This momentum-dependent mass term has recently been analyzed in a type I semi-Dirac material\cite{Marta2024}, where the authors show that  energy-dependent edge states appear only in one direction, localized on the upper or lower edge of a nanoribbon sample determined by their particle or hole character. The authors argue that the topological protection can be rigorously founded on the Zak phase of the one-dimensional reduction of the semi-Dirac Hamiltonian, which depends parametrically on one of the momentum degrees of freedom.
An interesting question is how these two mechanisms interplay to produce a distinct plasmon response in anisotropic type I semi-Dirac materials. 

Moreover, given the recent interest in separated two-layer systems of two-dimensional materials, it is important to determine the role of coupled anisotropic layers composed of semi-Dirac materials and how the anisotropy could impact the plasmonic response of the configuration. As was shown analytically and numerically by the authors in reference \cite{GiriPRB2022}, the time-reversal symmetry-broken Weyl semimetal thin film hosts a plasmon mode that results from collective antisymmetric charge oscillations between the two surfaces, which is in stark contrast to conventional two-dimensional bilayers as well as Dirac semimetals with Fermi arcs, which support antisymmetric acoustic modes along with a symmetric optical mode. The generic features of collective modes of spatially separated systems in two-dimensional plasma in solids was put forward in reference\cite{Sarma1981} where the authors showed that at long wavelengths, the spatial separation between the two charge components makes it possible for the acoustic branch to move out of the electron hole continua provided
it exceeds a critical value, which could be realized in a $\text{GaAs}-\text{Ga}_x\text{Al}_{1-x}\text{As}$ double quantum well. These spatially separated two-layer systems have recently been studied in graphene \cite{Brey2024} where the separation among layers is assumed to be large enough so that electron tunneling between layers can be
neglected. The layers are connected to metallic contacts that define their Fermi energies, and it is further assumed that both layers have the same density of carriers, which means the same plasmon dispersion $\omega_q$ when they are isolated.

The purpose of this work is twofold; first, we address the plasmon spectrum of long wavelength excitations in a single-layer of semi-Dirac material as the system undergoes the topological phase transitions, whenever the Dirac cone merging and/or band inversion regimes are reached separately; second, we study the role of the inherent anisotropic spectrum in a configuration of two layers of semi-Dirac materials as their main axes are rotated with respect to one another, showing the emergence of a second sharp direction-dependent plasmon mode. Interestingly, we find that the layers rotation angle affords a means to control the relative phase between the two plasmon modes, and this tunability could be exploited in  topological plasmonics and interferometric applications in next-generation optoelectronic devices.
The paper is organized as follows; in the next section, we present the semi-Dirac Hamiltonian and review some of its basic properties. Then, section \ref{section3} describes the plasmon response within the random phase approximation for the dielectric function. The results for single- and two-layer systems are analyzed in section \ref{section4}, whereas the discussion and conclusions of the work are given in section \ref{section5}. Finally, an appendix summarizing some relevant calculations is given at the end. 
\section{model}\label{section2}
For a monolayer of a semi-Dirac system, our effective model corresponds to the low-energy Hamiltonian description of spinless charged particles in a two-dimensional semi-Dirac material, which can be obtained as a $k\cdot p$  Hamiltonian approximation from a multi orbital Slater-Koster tight-binding approach\cite{Marta2024,PeraltaPRB2024} 
\begin{equation}\label{eq1}
H=\left(\frac{p_x^2}{2m^{*}}+\Delta\right)\sigma_x+vp_y\sigma_y+(M_0-M_1p^2)\sigma_z, 
\end{equation}
with $m^{*}$ being the effective mass and $v$ the Fermi velocity\cite{montambaux}. In addition, $(p_x,p_y)$ is the particle's momentum, within the long-wavelength approximation, and $\sigma_i$ with $i=\{x,y,z\}$, are Pauli matrices in pseudospin space, whereas the momentum-dependent mass term is given by $\Delta_p=M_0-M_1p^2$, where $M_0$ and $M_1$ appear in the minimal model\cite{BHZScience2006}. 
Thus, we have two parameters to control the topological properties of the system, the sign of $\Delta$ and the ratio $M_0/M_1$. For the minimal model of a semi-Dirac material $M_0=M_1=0$ such that for $\Delta<0$ two Dirac cones are located along the quadratic direction, at momenta $p_x=p_\pm=\pm\sqrt{-2m^{*}\Delta}$. For $\Delta=0$, these Dirac cones merge into two parabolic bands, touching at $p_x=0$, and the case $\Delta>0$ corresponds to the parabolic gapped bands. On the other hand, for $\Delta=0$, the scenario $M_0/M_1<0$ leads to trivial bands with a gap, followed by gapless touching bands when $M_0/M_1=0$ and inverted bands with a gap, when $M_0/M_1>0$\cite{Shen}. 
\begin{figure}
    \centering   \includegraphics[width=.675\linewidth]{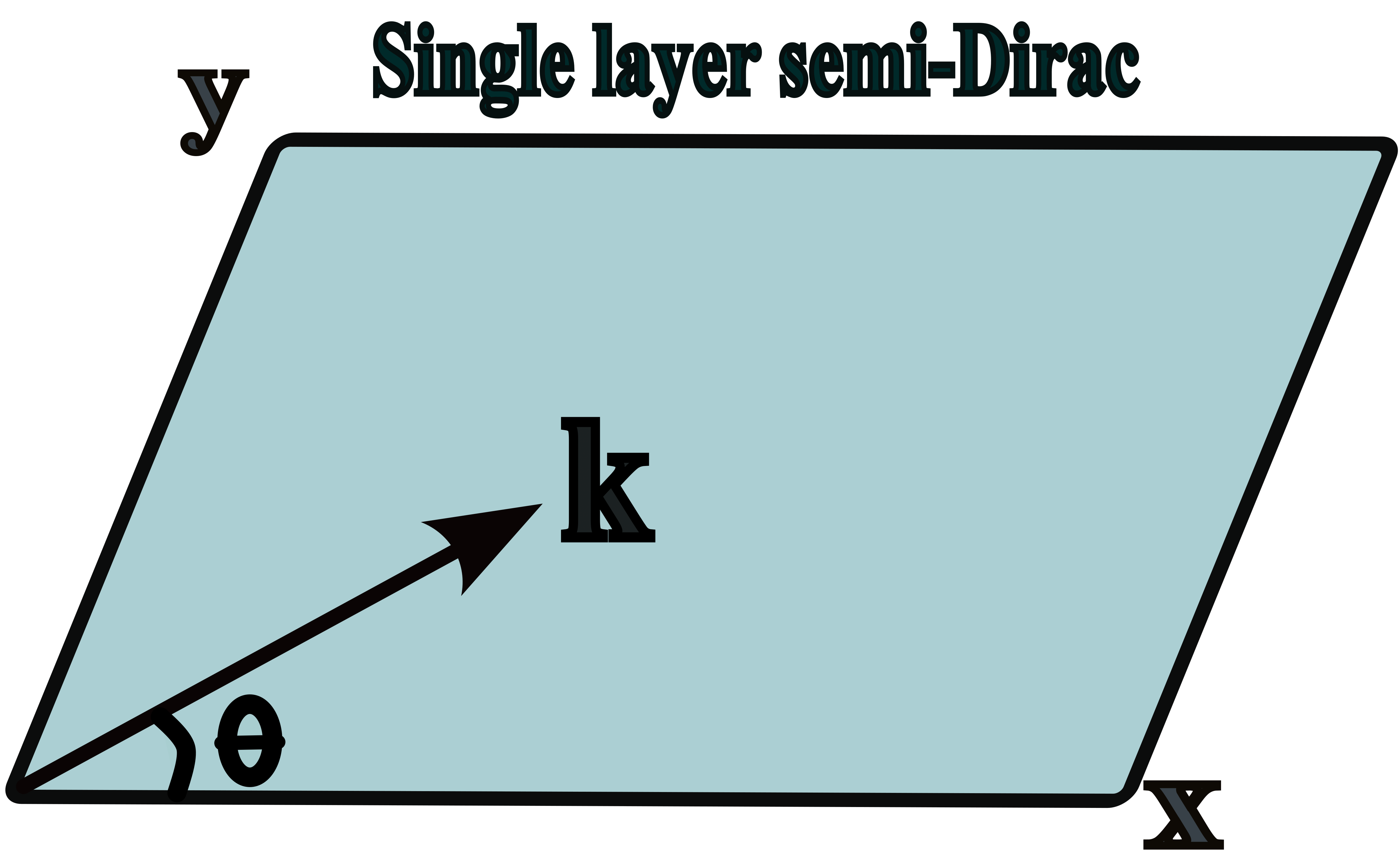} \includegraphics[width=1\linewidth]{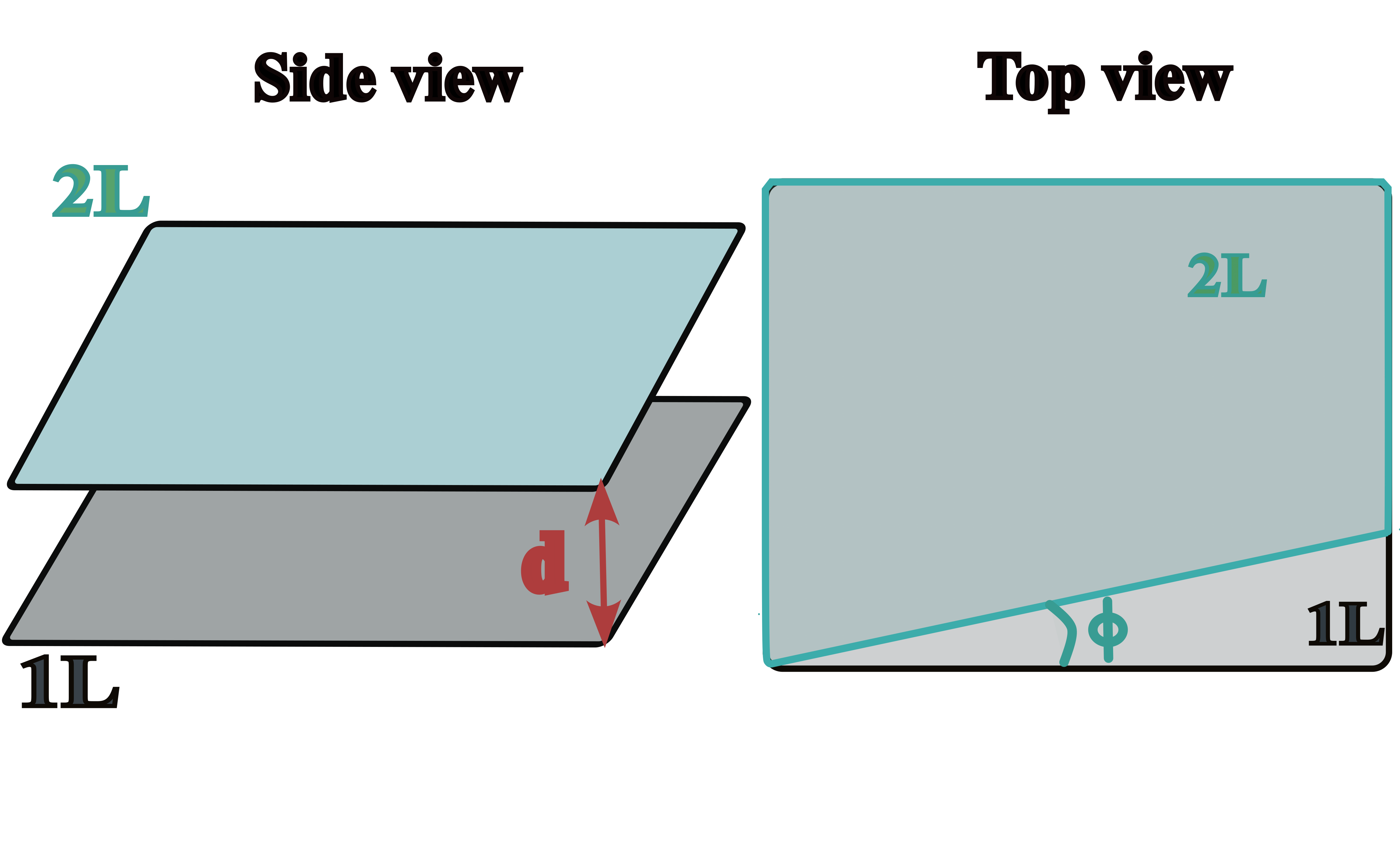}
    \caption{The upper figure shows the single-layer semi- Dirac configuration. The momentum ${\bf k}$ is shown along with its angle $\theta$ with respect to the horizontal $x$ axis. The lower left figure shows the side view of the composite two-layer system and the right figure shows a top view, where the relative angle $\phi$ gives the orientation of the top layer with respect to the lower one.}
    \label{fig:enter-label}
\end{figure}
In general, the anisotropy of the semi-Dirac material could also be inherited by the mass term, so we would have $M_{1x}\neq M_{1y}$, but this generic scenario would not qualitatively change our main results. For typical semi-Dirac material, the parameter values can be set as $\hbar v=0.65eV\AA$, $\hbar^2/(2m^{*})=0.75\text{eV}\AA^2$,and $\Delta=0.01\text{eV}$ which are reported in reference \cite{SahaPRB2016}, whereas the parameter values $M_0=0.09\text{eV}$ and $M_1=0.23\text{eV}\AA^{-2}$ are taken from reference \cite{Marta2024}.

In the following, it is convenient to write the Hamiltonian in dimensionless form, which is achieved by using $2m^{*}v^2$ as the energy scale. Thus, upon defining $k_x=p_x/2m^{*}v$, $k_y=p_y/2m^{*}v$, $\delta=\Delta/2m^{*}v^2$, $m_0=M_0/2m^{*}v^2$, and $m_1=2m^{*}M_1$, one gets the dimensionless Hamiltonian
\begin{equation}
H_d=(k_x^2+\delta)\sigma_x+k_y\sigma_y+\Delta_k\sigma_z,
\end{equation}
where $\Delta_k=m_0-m_1k^2$, with $k^2=k_x^2+k_y^2$.
The dimensionless energy spectrum is 
\begin{equation}
E_\pm({\bf k})=\pm\sqrt{(k_x^2+\delta)^2+ k_y^2+\Delta_k^2}=\pm E_k.
\end{equation}
This spectrum is shown in \figurename{2} along the quadratic (linear) direction $k_y=0$ ($k_x=0$). The continuous (dashed) lines correspond to vanishing (finite) momentum-dependent mass terms. In the left (right) panels the dashed lines correspond to $m_0/m\neq|\delta|$ ($m_0/m_1=|\delta|)$, considering $\delta=-1$. The dotted lines are for vanishing $\delta=m_0=m_1=0$, showing that for a conventional type I semi-Dirac monolayer sample, the quadratic (linear) bands touch at $k_x=0$ ($k_y=0$), while for finite $\delta$ and vanishing mass terms, the continuous lines show the generation of Dirac cones (parabolic gapped bands) along the $k_x$ ($k_y$) direction in momentum space. These Dirac cones along the $k_x$ emerge only for $\delta<0$ and are located at $k_\pm=\pm\sqrt{-\delta}$.  Interestingly, when the mass terms are finite (dashed lines)
$m_0/m_1 =1$, the mass term can either break or preserve
the Dirac cones for these sets of parameters. This condition is also explored along the linear dispersion direction $k_y=0$. However, along this direction in the momentum space the spectrum is always gapped for finite values of  $\delta$ and/or the momentum-dependent mass terms $m_0,m_1\neq0$.
\begin{figure}[h]
\includegraphics[height=2.5cm]{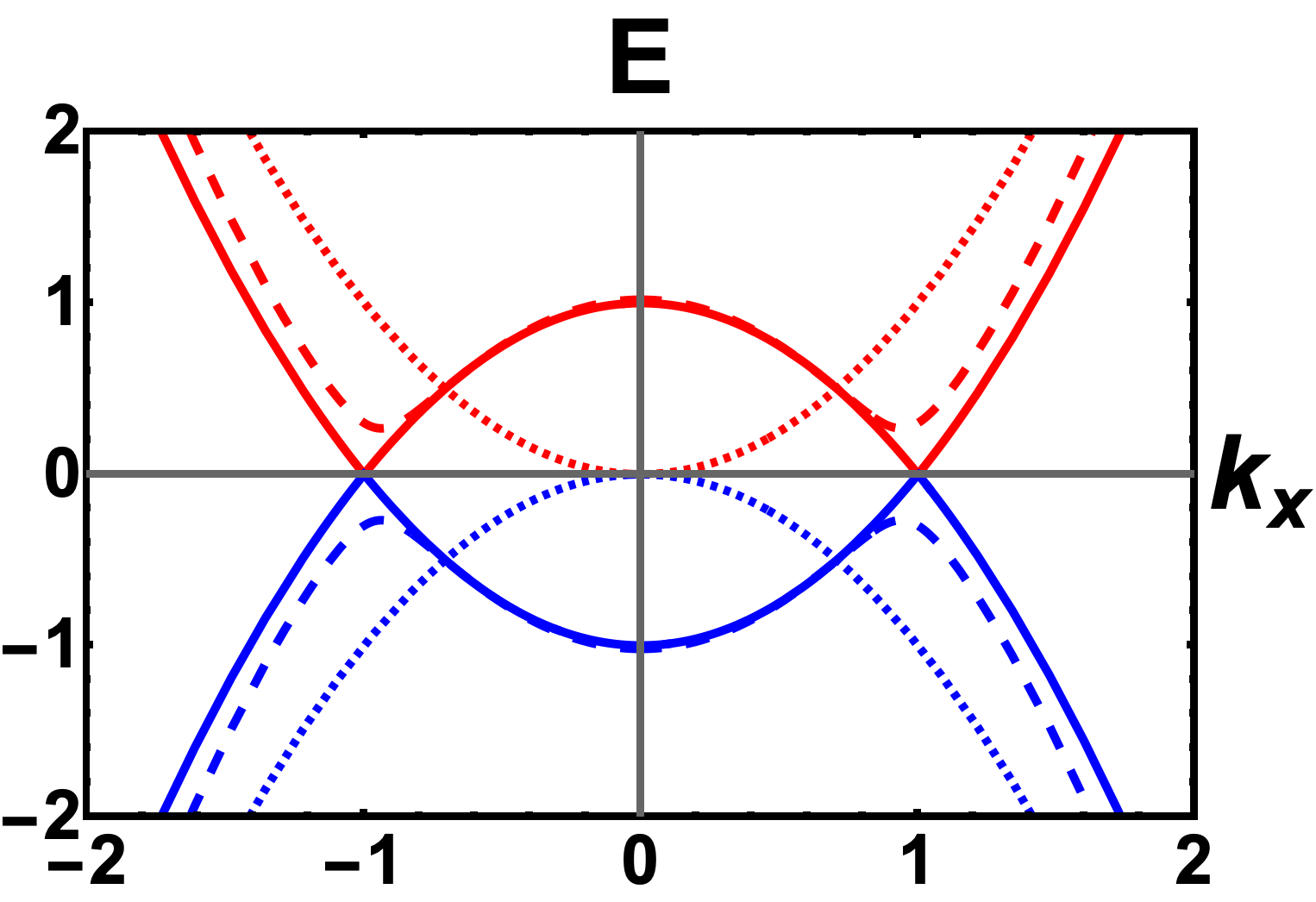}
\includegraphics[height=2.5cm]{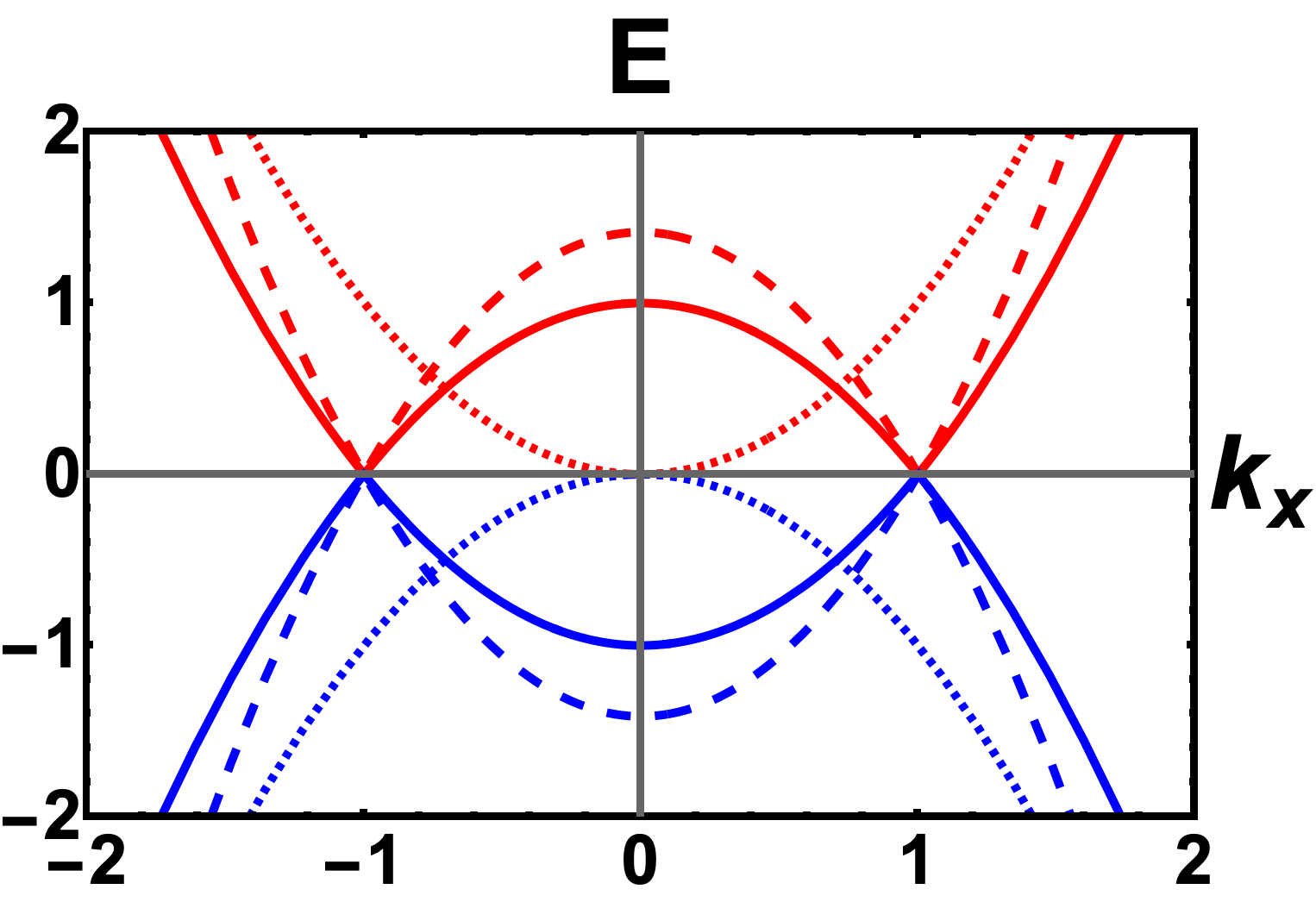}
\includegraphics[height=2.5cm]{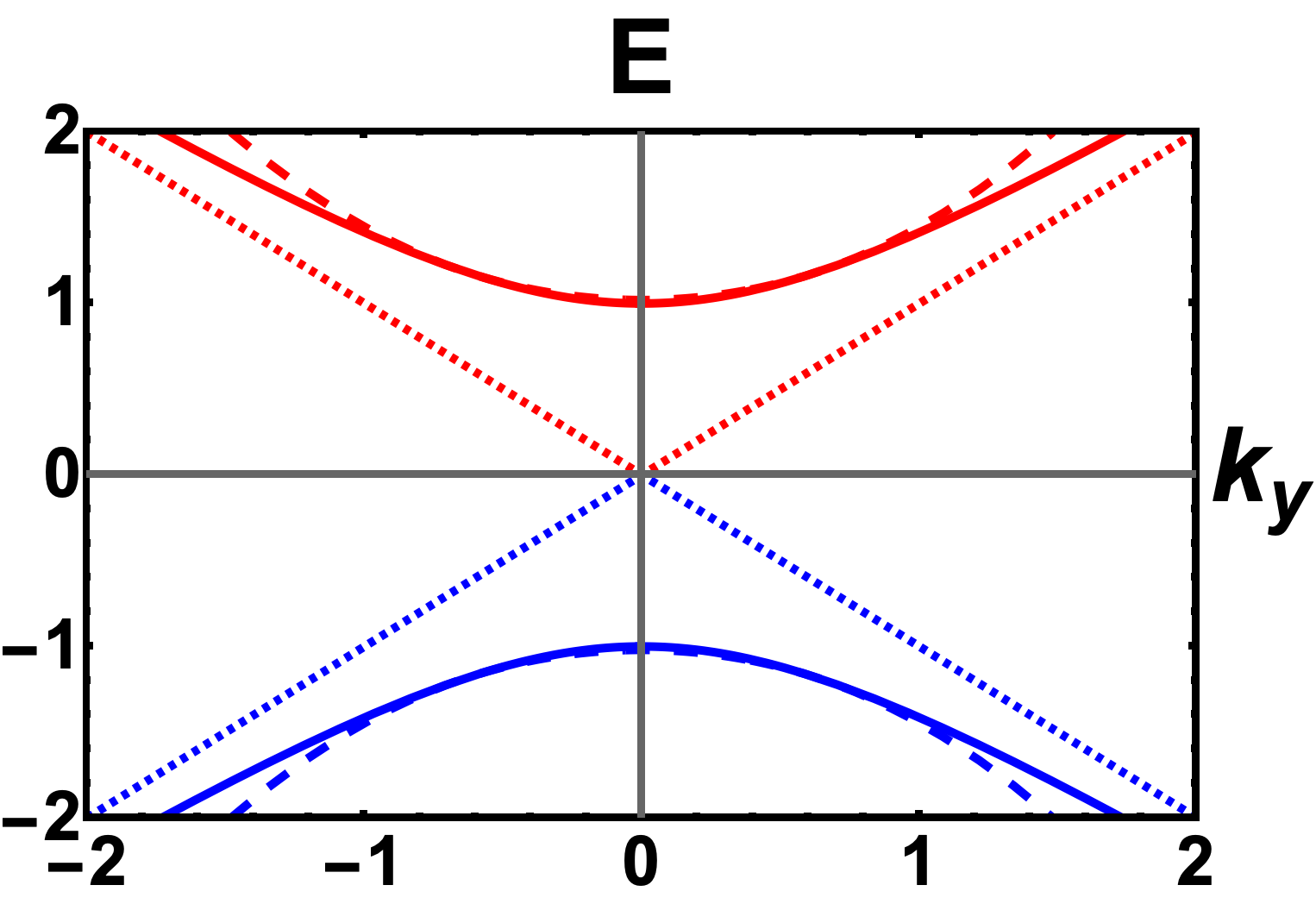}
\includegraphics[height=2.5cm]{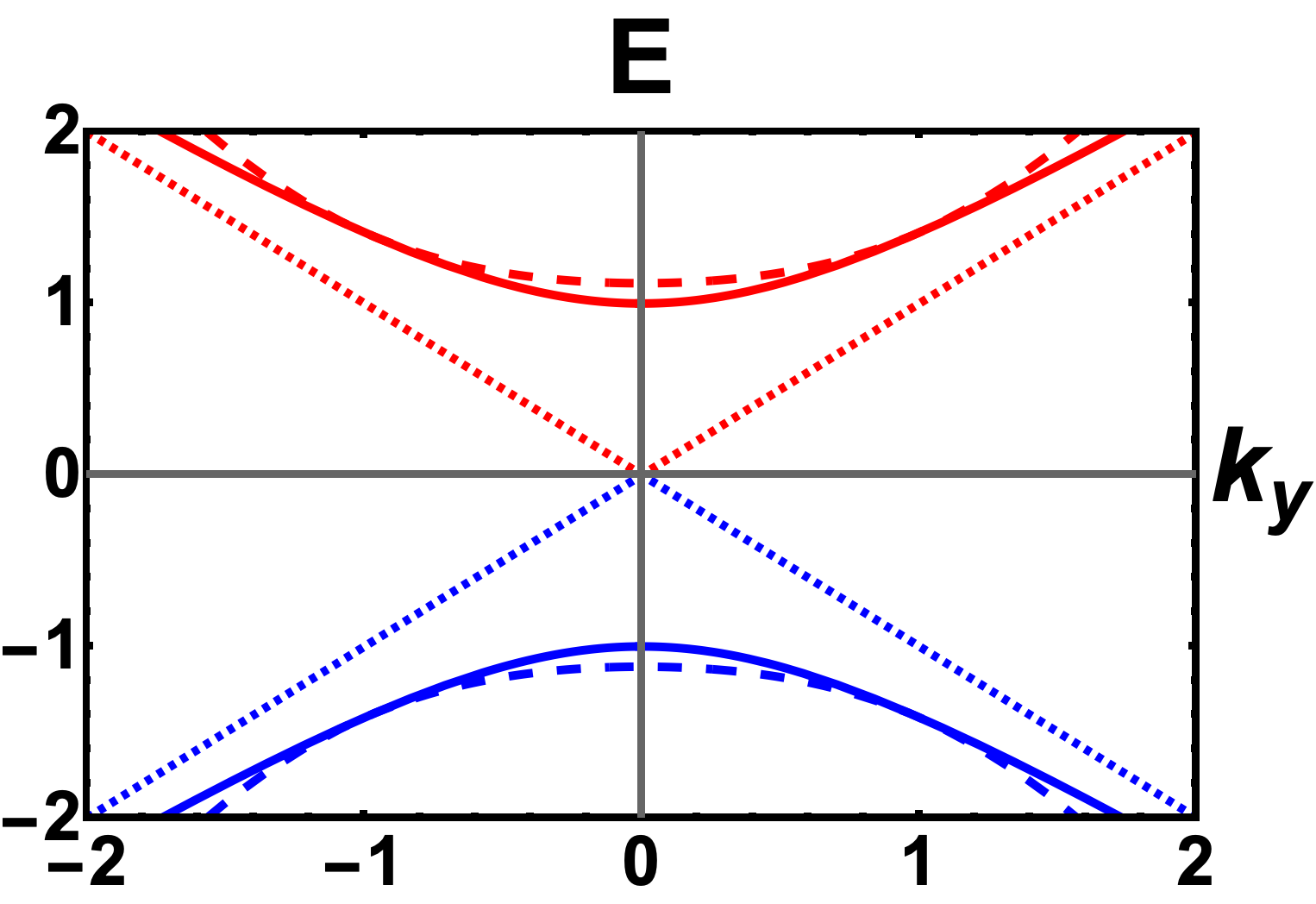}
\caption{Energy-momentum dispersions along the quadratic (linear) direction $k_y=0$ ($k_x=0$) are shown in the upper (lower) panels. The thin dotted lines correspond to $\delta=m_0=m_1=0$, while the continuous lines correspond to  $\delta=-1$, $m_0=m_1=0$. The dashed lines in the left (right) panels  correspond to $m_0/m_1=|\delta|$ ($m_0/m_1\neq|\delta|)$, for $\delta=-1$.}
\end{figure}
 The corresponding eigenstates read as follows,
\begin{equation}\label{eq4}
|\boldsymbol{k}s\rangle=\frac{1}{\sqrt{2E_k}}
\begin{pmatrix}
\sqrt{E_k+s \Delta_k}\\
s e^{i\varphi_k}\sqrt{E_k-s\Delta_k},
\end{pmatrix}
\end{equation}
where $\tan\varphi_k=k_y/ (k_x^2+\delta)$ and $s=\pm1$. These eigenstates will be used for the calculation of the plasmon spectrum in the next section.
\section{Plasmon response in monolayer semi-Dirac material}\label{section3}
\subsection{Single layer system}
\noindent The plasmon spectrum in two-dimensional anisotropic materials such as the anisotropic lattice of phosphorene determines different propagation of plasmons along the armchair and zigzag directions. Black phosphorus is also a suitable material for ultra-fast plasmonics, for which the active plasmon response can be initiated by photoexcitation with femtosecond pulses\cite{Agarwal2018}. In this section, we consider the Random Phase Approximation (RPA) to describe the plasmon spectrum of a monolayer type I semi-Dirac material in the presence of momentum-dependent mass terms $m_0/m_1$ and the Dirac merging parameter $\delta<0$. 
To obtain the plasmon spectrum for the monolayer system, we calculate the dielectric function,
\begin{equation}
\epsilon(\boldsymbol{q},\omega)=1-V(\boldsymbol{q})\Pi(\boldsymbol{q},\omega)
\end{equation}
where $V(q)=\frac{e^2}{2\epsilon_0\epsilon_rq}$ is the Fourier transform of the Coulomb interaction $V(r)=\frac{e^2}{4\pi\epsilon_0\epsilon_rr}$ in two dimensions, with $\epsilon_0$ the vacuum permittivity and $\epsilon_r$ the background dielectric constant. The plasmon modes are obtained by looking for zeros of the dielectric function. In addition, the non-interacting polarizability function $\Pi(\boldsymbol{q},\omega)$, within the linear response regime, is defined as
\begin{widetext}
\begin{equation}
\Pi(\boldsymbol{q},\omega)=\frac{1}{4 \pi^2} \displaystyle\int d^2k \sum_{s,s'=\pm1}\frac{n^{s}(\boldsymbol{k})-n^{s'}(\boldsymbol{k'})}{\hbar \omega+E_{s}(\boldsymbol{k})-E_{s'}(\boldsymbol{k'})+i0} F^{ss'}(\boldsymbol{k},\boldsymbol{k'}),  
\end{equation}    
\end{widetext}
with $\boldsymbol{k'=k+q}$, while $\omega$ and $\boldsymbol{q}$ are the plasmon frequency and momentum, respectively. Here $n^s(\boldsymbol{k})=[e^{(sE_k-\mu)/k_BT}+1]^{-1}$  is the Fermi-Dirac distribution function, which at zero temperature can be replaced with the Heaviside step function $\Theta(sE_k-E_F)$, with $E_F$ being the Fermi energy. In addition, the quantity $F^{ss'}(\boldsymbol{k},\boldsymbol{k'})=|\langle \boldsymbol{k}s|\boldsymbol{k'}s' \rangle|^2$ gives the overlap of the Hamiltonian eigenstates in eq. (\ref{eq4}), evaluated at the different momenta and pseudospin values, describing the weighting of the optical transitions in the intraband $s'=s$ and interband $s'\neq s$ processes. 
\subsection{Two-layer plasmon spectrum}
For a system composed of two spatially separated layers of semi-Dirac material, contacted through a dielectric material, we assume that their separation is large
enough so that electron tunneling between layers can be
neglected, yet the Coulomb interaction between layers is strong enough to modulate the polarization effects \cite{Brey2024}. In addition, the layers are connected to metallic contacts that define their Fermi energies. We also assume that both layers have the same density of carriers and, therefore, the same plasmon dispersion $\omega$ when they are isolated. 
If the polarization function for each layer is $\Pi_{1,2}(\vec{q},\omega)$, where the subindex labels the semi-Dirac layer, which can be rotated with respect to each other as shown schematically in \figurename{1}, with relative angle $\phi$. To account for this relative layer rotation, we parametrize the momenta in the upper layer as
\begin{eqnarray*}
k_{x_2}&=&k_{x_1}\cos\phi+k_{y_1}\sin\phi\\
k_{y_2}&=&-k_{x_1}\sin\phi+k_{y_1}\cos\phi.
\end{eqnarray*}
To account for the bare intra- and interlayer Coulomb interactions we use the results from references \cite{GiriPRB2022,bilayer1,bilayer2,bilayer3,bilayer4,bilayer5,bilayer6,bilayer7,bilayer8,bilayer9,bilayer10} where the dielectric function for the bilayer configuration is given by the matrix 
\begin{equation}\label{bilayer-epsilon}
\epsilon(\boldsymbol{q},\omega)=
\left(
\begin{array}{cc}
 1-V_{11}\Pi_1    &V_{12}\Pi_1  \\
 -V_{21}\Pi_2    & 1-V_{22}\Pi_2
\end{array}  
\right),
\end{equation}
where $V_{ij}\equiv V_{ij}(\vec{q},\omega)$, with $i,j=1,2$. Since the two layers are equivalent, one gets $V_{11}= V_{22}=\alpha/q$, whereas for the off-diagonal contributions one needs to take into account the layer separation, which is modeled with the effective interaction term  $V_{12}=V_{21}=\alpha  e^{-qd}/q$,  
where $d$ is the layer's separation.  Unless otherwise stated, the distance between the two layers would be fixed at $d=5$, in units of $\hbar/m^{*}v$. In addition, $\alpha = 2\pi e^2/\epsilon$, with the effective dielectric constant of the bulk separating the semi-Dirac layers $\epsilon=\epsilon_0\epsilon_r$. For this two-layer configuration, the sharp plasmon modes can be found from the condition of the vanishing of the determinant of the dielectric matrix eq. (\ref{bilayer-epsilon}). Double‐layer two‐dimensional systems are known to host two distinct plasmon branches. In the long‐wavelength limit, the higher‐energy optical mode follows a $\sqrt{q}$ dispersion, whereas the lower‐energy acoustic mode scales linearly with $q$. In the next section, we discuss the main salient features of single- and two-layer systems, contrasting their distinct features. Unless otherwise stated, we fix $\mu=1.05$ in units of $2m^{*}v^2$.
\begin{figure}[ht]
\includegraphics[width=0.48\textwidth]{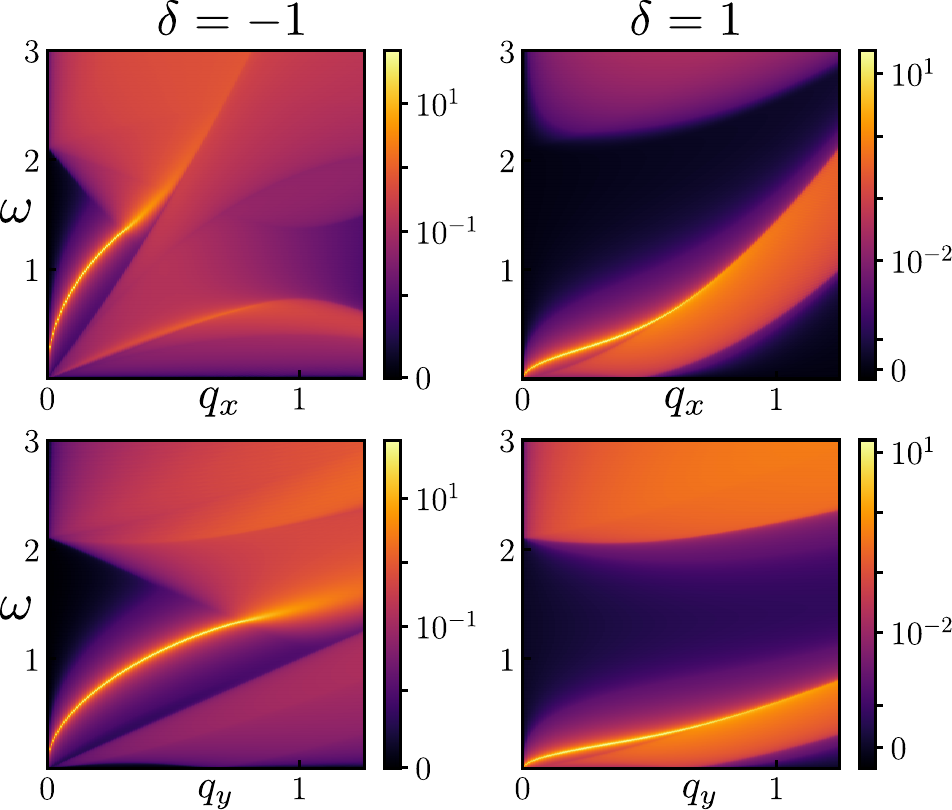}
\caption{The inverse dielectric function($-\Im \frac{1}{\epsilon(\boldsymbol{q},\omega)}$) for a single layer of semi-Dirac material corresponding to $\delta=-1$ (left panels) and $\delta=1$ (right panels), as a function of $\omega$ along the two main axis directions in momentum space. The bright yellow curve marks the peak of $-\Im \frac{1}{\epsilon(\boldsymbol{q},\omega)}$ indicating sharp plasmon resonances. We have fixed  $m_0=m_1=0.0$.}
\end{figure}\label{plasmon_delta1_deltam1}
\section{Results}\label{section4}
For the single-layer semi-Dirac system with $m_0=m_1=0$, we show in \figurename{3} the inverse of the imaginary part of the dielectric function  corresponding to $\delta=-1$ (left
panel) and $\delta=1$ (right panel), as a function of $\omega$ along the
two main axis directions in momentum space. The bright yellow
curve marks the peak of  $-\Im\frac{1}{\epsilon(\boldsymbol{q},\omega)}$, indicating sharp plasmon resonances. 
\begin{figure}[h]
\includegraphics[width=0.5\textwidth]{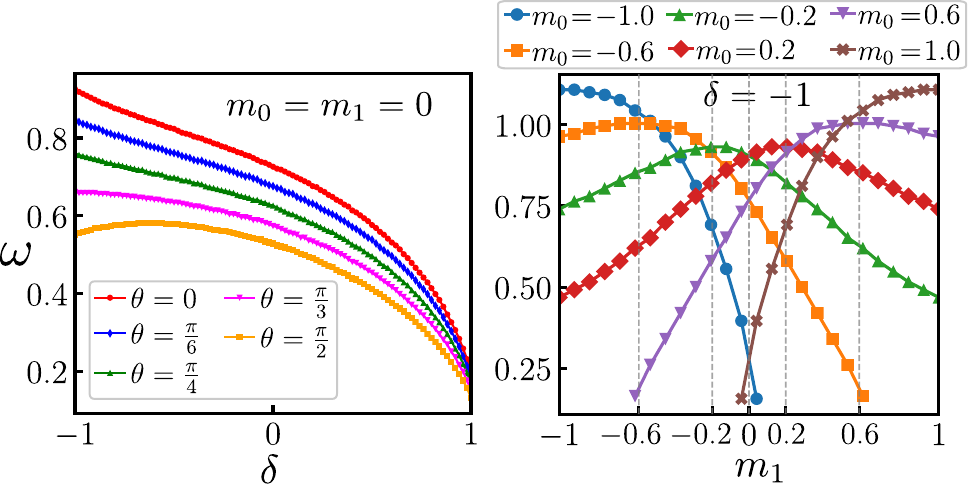}
\caption{Parametric plots of the plasmon frequency response for a single layer of semi-Dirac material as a function of $\delta$ (left panel) and as a function of $m_1$, corresponding to different $m_0$ (right panel). We have fixed the dimensionless plasmon momentum at $q=0.2$. For the left plot, we choose the relative angle between $q_x$ and $q_y$ to be $\theta=0,\pi/6, \pi/4, \pi/3, \pi/2 $, and for the right plot we have set $\theta=0$, which highlights the role of Dirac cone merging at finite mass terms. The vertical lines are a guide to the eye showing the respective maxima in frequency for $m_0/m_1=|\delta|.$}
\end{figure}\label{plasmon_delta_m0_m1}
Given the anisotropy of the energy spectrum, we separately study the plasmon response along the two momentum directions. We first consider  the case $\delta<0$, along the quadratic direction $q_x$
that, at sufficiently large momenta, shows a dominant linear dependence and at lower momenta,  the dominant optical plasmon mode behavior $\sqrt{q_x}$ is verified. These two behaviors are observed in the upper left figure of \figurename{3} where a larger frequency response of the plasmon is clearly seen. Thus, the plasmon response could serve as an optical marker for the underlying topological properties associated with the Dirac cone merging phase. 
In addition, the characteristics of the dominant optical mode $\omega\propto\sqrt{q_y}$ are clearly seen in the lower plots of \figurename{3}, where one notices, however, that the transition to linear behavior is not observed. This is a consequence of the absence of topological features associated with Dirac cones along this direction, which occurs whenever $\delta$ is finite, and is explicitly shown in the two lower plots of \figurename{2}.
\noindent On the other hand,  the lower plots of \figurename{3} show that although along the linear dispersion direction $q_y$  the $\delta<0$ parameter regime shows a larger frequency range than $\delta>0$, the plasmon resonance is damped at lower frequency values than those observed along the $q_x$ direction. 

To better understand the emergent physical characteristics, as $\delta$ continuously varies from negative to positive values, on the left plot of \figurename{4} we show the parametric frequency response along different momenta angles $\theta$, where $\tan\theta=q_y/q_x$. The observed response is that at $\delta<0$ the plasmon frequency value is larger and is dramatically reduced as we approach $\delta=+1$, showing that the result is a generic feature of the Dirac cone merging scenario, which occurs as long as $\theta<\pi/2$. The highest plasmon frequency is precisely along the main axis $q_x$ ($\theta=0$), which is shown by the red curve in the plot of the left of \figurename{4}. In addition, the role of the band inversion mass terms is presented in the right panel of \figurename{4}, where a parametric plot of the plasmon frequency  along $q_x$ ($\theta=0$), is given as a function of the band inversion mass term $m_1$. This is parametrized by selected values of the band inversion mass term $m_0$, for the interesting topological Dirac cone regime $\delta=-1$. The generic trend is that as $m_0/m_1\rightarrow |\delta|$ the plasmon frequency tends to a maximum value, which is consequent to the emergence of Dirac cones for finite band inversion mass terms shown by the dashed lines in the upper right plot of \figurename{2}. This in turn highlights the relevance of Dirac cone emergence at finite-band inversion mass terms, indicating the cooperative effects among these two mechanisms for generation of nontrivial topological phases.
\begin{figure}[h]
\includegraphics[width=0.48\textwidth]{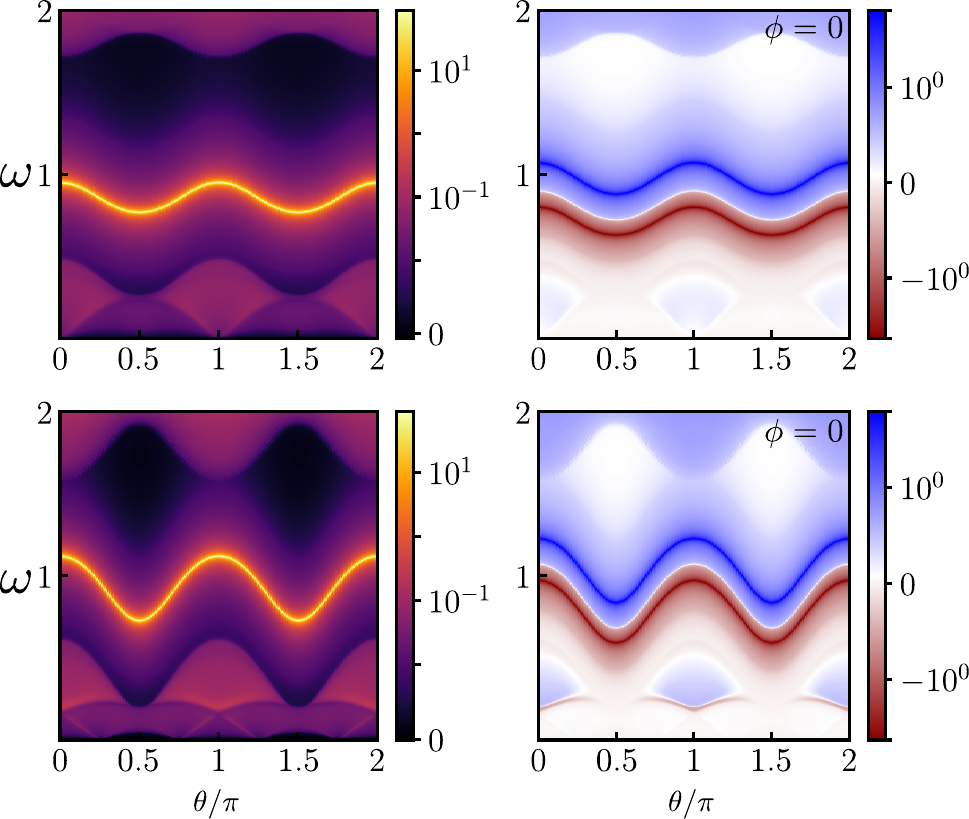}
\caption{Angular dependence of the inverse dielectric function($-\Im \frac{1}{\epsilon(\theta,\omega)}$) for single (left panels) and double layer (right panels) at fixed $q=0.2$. In the left panels, the bright yellow curve marks the peak of $-\Im \frac{1}{\epsilon(\theta,\omega)}$, while in the right panels, the dark red (and blue) curves indicate coupled plasmon resonances, using $-\Im [{\rm det}[1/\epsilon(\theta,\omega)]]$. The upper panels correspond to vanishing merging parameter $\delta=0$ and band inverted condition $m_0/m_1>0$, with $m_0=0.2$ and $m_1=0.5$. The lower panels correspond to the Dirac cone condition $\delta=-1$ and vanishing mass terms $m_0=m_1=0$. For the double layer system we have set $\phi=0$.} 
\end{figure}\label{plasmon1}
We can gain further insight by analyzing the plasmon spectrum of the two-layer system schematically depicted in  \figurename{1}. For this purpose, in \figurename{5} we show the angular dependence of the imaginary component of the inverse dielectric function for single- and double-layer systems, where we have set the effective dimensionless plasmon momentum at $q=0.2$. The plasmon spectra for the two-layer configuration are labeled by the relative layer angle $\phi=0$, which indicates that within this configuration, the axes of the two-layers are perfectly aligned. The upper plots correspond to the vanishing merging
parameter $\delta=0$ and finite value $m_0/m_1 > 0$, with $m_0 = 0.2$ and $m_1 = 0.5$, while the lower plots correspond to the finite Dirac cone merging condition ($\delta=-1$), and vanishing mass terms $m_0=m_1=0$. To also highlight the different nature of the plasmon response for the single- and double-layer configurations, we use different color maps.  In the left panel, the bright yellow curve marks the peak of -$\Im 1/\epsilon(\theta,\omega)$, while in the right panel, the dark red and blue
curves identify these maxima, both indicating sharp plasmon resonances. 
\begin{figure}[h]
\includegraphics[width=0.5\textwidth]{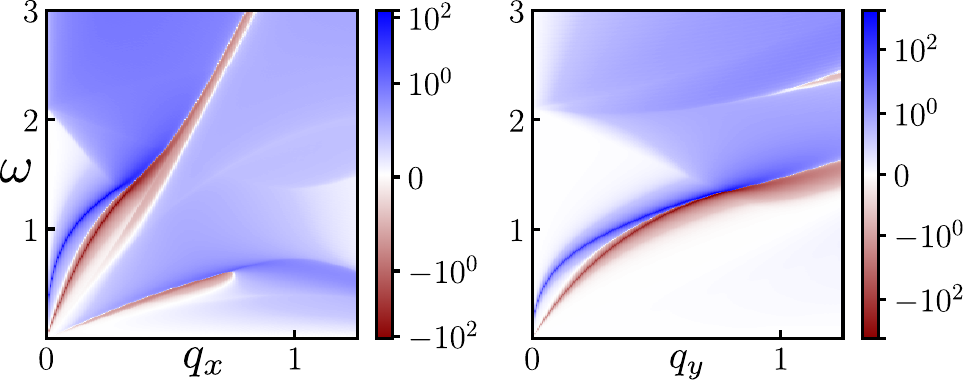}
\caption{The inverse dielectric function($-\Im [{\rm det}[\frac{1}{\epsilon(\boldsymbol{q},\omega)}]]$) for a double layer system as a function of $\omega$ along the two main axis directions in momentum space. The chosen parameters are $m_0=m_1=0.0$, and $\delta=-1$. The dark red (and blue) curves identify the maxima of $-\Im [{\rm det}[\frac{1}{\epsilon(\boldsymbol{q},\omega)}]]$, indicating sharp plasmon resonances. We consider that both layers are aligned.}
\end{figure}\label{plasmon4}
\noindent One sharp plasmon mode is observed for the monolayer, whereas an additional plasmon mode emerges in the two-layer system. For this configuration of a perfectly aligned two-layer system, the plasmon modes are in phase with respect to each other. Interestingly, since the momentum-dependent mass term is isotropic, for the scenario with inverted bands $m_0/m_1>0$ and $\delta=0$ the underlying anisotropic nature of the semi-Dirac spectrum shows a less pronounced angular variation in the frequency separation among the maximum and minimum values which are achieved along the linear and quadratic directions in momentum space, respectively. On the other hand, the lower plots of \figurename{5} show that, at finite $\delta<0$ and vanishing $m_0=m_1=0$, the anisotropic nature of the spectrum takes over and leads to a stronger angular difference in the plasmon response, with a larger characteristic plasmon frequency along the quadratic direction in momentum space. This once again highlights the role of the presence of the Dirac cones, which affects the optical response of the material. As expected, a $\pi$ periodicity associated with the inversion transformation $\theta\rightarrow \theta+\pi$ (i.e. $\vec{q}\rightarrow -\vec{q})$ is clearly seen in the plasmon frequency response when the two Dirac cones are present, as is explicitly shown in the lower plots of Fig. \figurename{5}. We also take into account that the dimensionless frequency $\omega$ is of the order of the energy separation in the interband transitions, which upon restoring the energy units, we obtain for the single-layer system along the quadratic direction $\theta=0$ a plasmon frequency $\hbar\Omega\approx(2m^{*}v^2$, which for the finite momentum mass ratio $m_0/m_1>0$ (top left plot) gives $\hbar\omega\approx0.51eV$, while for the vanishing mass ratio and finite $\delta=-1$ (lower left plot), one obtains the larger value $\hbar\omega\approx0.68eV$.
\begin{figure}[h]
\includegraphics[width=0.5\textwidth]{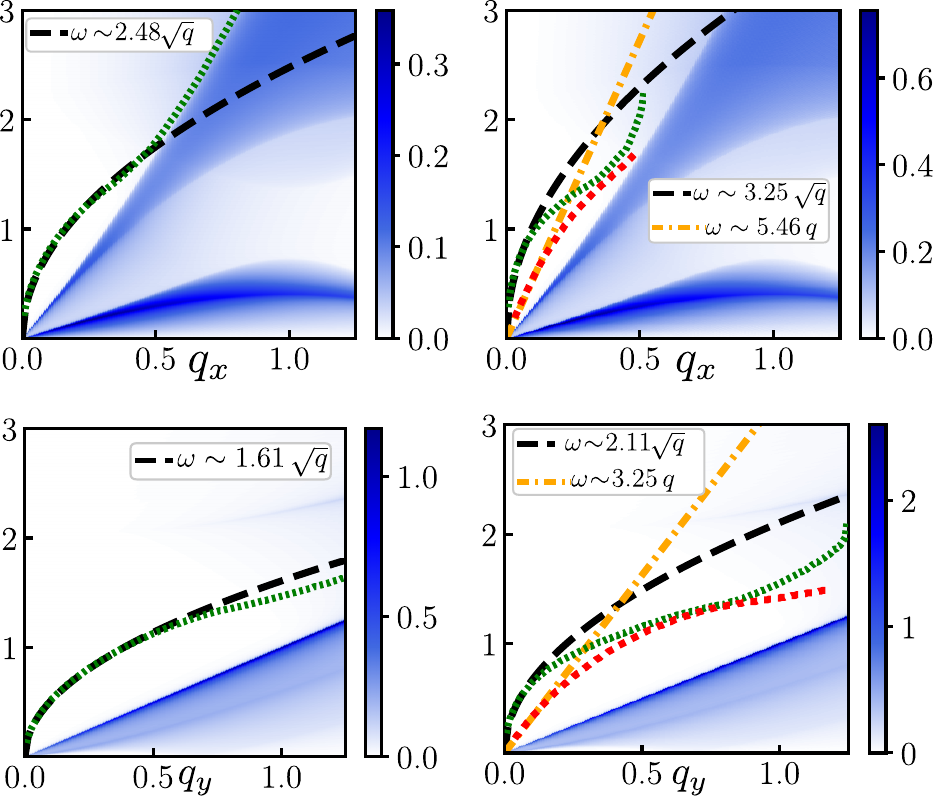}
\caption{The color bar represents the non-interacting polarizability function, $\Pi(\boldsymbol{q},\omega)$ as a function of $\omega$ along the two main axis directions in momentum space for single (double) layer shown in the left (right) panels. The chosen parameters are $m_0=m_1=0$, and $\delta=-1$. The black dashed line is a $\sqrt{q}$ fitting valid at momenta $|q|\lesssim0.5$, while the green (small dotted) line shows the actual plasmon response, computed from the zeros of $\epsilon(\boldsymbol{q},\omega)$. 
For the bilayer system with $d=5$, two distinct plasmon branches emerge: one following a $\sqrt{q}$ law (black dashed fit) and the other displaying a linear $q$ dependence (yellow dash-dotted fit), both valid up to $|q|\lesssim0.1$. The actual plasmon modes are shown by the green (square‐root branch) and red (linear branch) dotted lines.}
\end{figure}\label{plasmonfitting}
For the two-layer semi-Dirac system we show in \figurename{6} the plasmon frequency response for the aligned configuration along the two main directions in momentum space for $m_0=m_1=0$ and $\delta=-1$. The second plasmon mode is observed along both directions, and the trend of a larger frequency response along $q_x$ is preserved in the two-layer configuration. 
\figurename{7} shows a color map depicting the non-interacting polarizability $\Pi(\boldsymbol{q},\omega)$ as a function of frequency and momentum direction, onto which we overlay the plasmon dispersions extracted from the zeros of the dielectric function $\epsilon(\boldsymbol{q},\omega)$. For the monolayer (left panel), the plasmon mode (green dotted line) follows the $\sqrt{q}$ dispersion (black dashed) for $q\lesssim 0.5$ along both the $q_x$ and $q_y$ directions; beyond this momentum scale, a significant change in the behavior of the plasmon mode occurs along $q_x$, signaling direction-dependent dispersion of plasmons at higher wave vectors. In contrast, the two-layer system (right panel) supports two collective modes: one plasmon mode retains the behavior of $\sqrt{q}$ (black dashed) called optical mode, while the other mode exhibits a linear dependence of $q$ (yellow dashed-dotted) called acoustic mode, both valid for $q\lesssim 0.1$. The fact that the two-layer plasmon branches veer away from their ideal $\sqrt{q}$ and linear forms at much smaller $q$ directly highlights the dominant role of the interlayer Coulomb kernel in coupled systems. Physically, the coupling between layers modifies the effective Coulomb potential from its single-sheet $1/q$ form to a momentum‐dependent kernel proportional to $(1- e^{-qd})/q$.  
\begin{figure}
\includegraphics[width=0.48\textwidth]{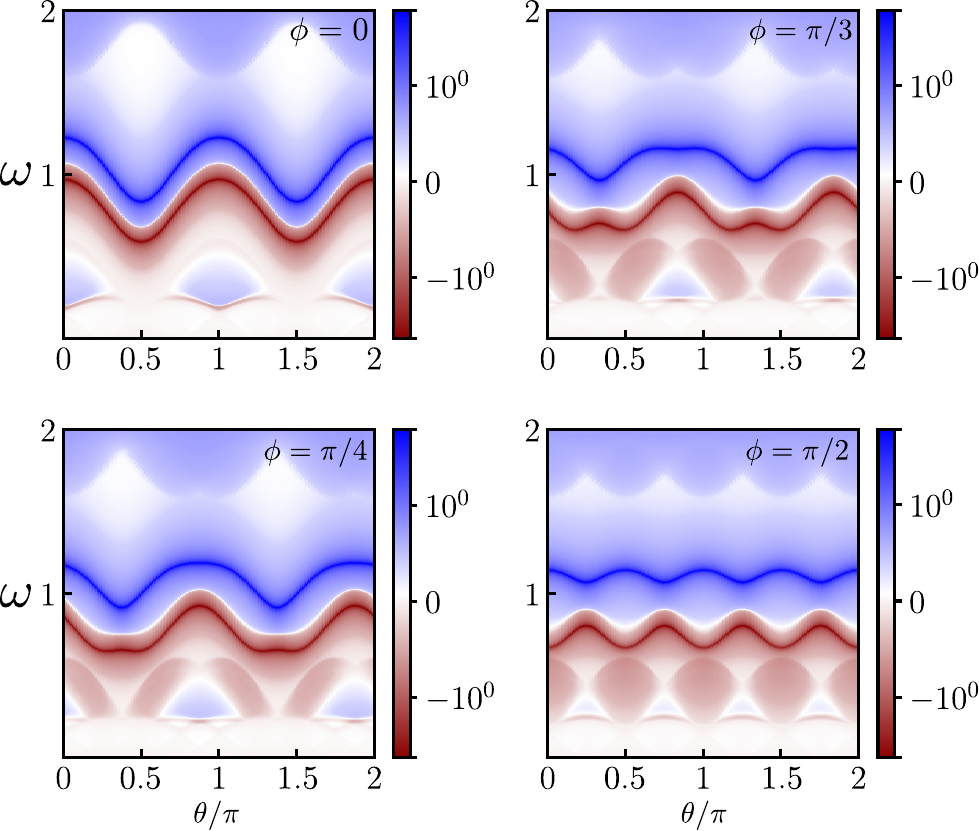}
\caption{The angular dependence of inverse dielectric function($-\Im \frac{1}{\epsilon(\theta,\omega)}$) for a double layer when the upper layer main axes are rotated by $\phi=0,\pi/3, \pi/4,\pi/2$ with respect to the positive x-axis in the momentum space in the lower layer. The dark red (and blue) curves correspond to the peak of $-\Im \frac{1}{\epsilon(\theta,\omega)}$,  indicating sharp plasmon resonances. The chosen parameters are $m_0=m_1=0$, and $\delta=-1$ and the momentum magnitude is set to $q=0.2$. }
\end{figure}\label{plasmon5}
This exponential factor dramatically reshapes the dispersion of the plasmon modes. In practical terms, this means that two-layer structures offer a tunable platform in which one can engineer when, and over what momentum range, the optical and acoustic plasmons deviate from simple analytic laws, opening new possibilities for mid-infrared and Terahertz modulators, where controlled dispersion is essential.

The effects of relative layer rotation are encoded in \figurename{8}, where we show the angular dependence of the inverse dielectric function ($-\Im [{\rm det}[\frac{1}{\epsilon(\theta,\omega)}]]$) when the main axes of the upper layer are rotated by $\phi=0,\pi/3, \pi/4,\pi/2$ with respect to the positive x axis of the lower layer. The dark red and blue curves correspond to the peak of $-\Im [{\rm det}[\frac{1}{\epsilon(\theta,\omega)}]]$,  indicating sharp plasmon resonances. The chosen parameters are $m_0=m_1=0$, and $\delta=-1$, and the magnitude of the momentum is set to $q=0.2$. Thus, it is found that the relative angle impacts the relative phase between the two plasmon modes, which are in phase for aligned layers and transition to completely out of phase regime for $\phi=\pi/2$, i.e., when the upper layer's quadratic and linear momentum dispersions are aligned with the corresponding linear and quadratic directions in the lower layer. 

Furthermore, we analyze the eigenvalues and eigenvectors of the dielectric matrix $\epsilon(\omega,{\bf q})$ in the plasmon resonance condition, defined by ${\rm det[\epsilon(\omega,{\bf q})]}=0$. At this point, an eigenvalue vanishes, and the corresponding eigenvector characterizes the charge distribution between the two layers, revealing the symmetry of the collective mode. 
For perfectly aligned layers ($\phi=0$), along $\theta=0$ and at $|q|=0.2$, the optical mode with higher energy (dispersing as $\sqrt{q}$) has an eigenvector $(1/\sqrt{2},1/\sqrt{2})$, indicating a symmetric charge oscillation across the layers, while the acoustic mode (linear in $q$) has $(1/\sqrt{2},-1/\sqrt{2})$, corresponding to an antisymmetric configuration \cite{bilayer1,bilayer2,bilayer3,bilayer4,bilayer5,bilayer6,bilayer7,bilayer8,bilayer9,bilayer10}. For $\phi=\pi/2$, these eigenvectors deform approximately to $(0.98,0.198)$  for the optical mode and $(-0.396,0.918)$ for the acoustic mode. This shows that while the ideal symmetry is broken due to relative orientation, the optical and acoustic modes retain their predominantly symmetric and antisymmetric character, respectively, with layer-dependent weights modulated by the layer's geometric axis relative orientation.
 A similar trend is observed along $\theta=\pi/2$: in the aligned case, the eigenvectors retain their ideal symmetric or antisymmetric charge oscillation. However, for $\phi=\pi/2$, one finds that the eigenvectors of optical and acoustic modes are $(0.198,0.98)$ and $(0.918,-0.396)$ respectively. In particular, along $\theta=\pi/2$, the eigenvectors for the $\phi=\pi/2$ configuration appear to be approximately exchanged compared to those in $\theta=0$. This exchange reflects the underlying symmetry of the system and indicates that the layer participation in each mode is not fixed, but depends sensitively on the relative orientation between the two layers and the momentum direction. Rotated configurations $\phi=\pi/4,\pi/3$ show additional phase relations, showing that this two-layer semi-Dirac system could be suitable to exploit the anisotropic properties in future experimental implementations, for instance, in interferometric configurations.
\section{conclusions}\label{section5}
In this work, we have investigated the modulation of the plasmonic response in semi-Dirac materials, focusing on the influence of the parameter $\delta$, which governs the topological phase transition associated with the generation and merging of Dirac cones and its interplay with momentum-dependent band inversion mass terms $m_0$ and $m_1$, whose ratio $m_0/m_1$ controls the onset of topological transitions via band inversion, akin to those observed in topological insulators.
We analyze and compared the key features of both single-layer and bilayer semi-Dirac configurations. In the single-layer case, we found that along the momentum direction where Dirac cone generation occurs, the presence of Dirac cones enhances the range of plasmon frequencies associated with collective excitations. Interestingly, even though finite values of the band inversion terms typically lead to a gapped spectrum and eliminate the Dirac cones, we observe that the plasmon frequency reaches a maximum when the condition $m_0/m_1 = -\delta$ is satisfied for $\delta < 0$. Additionally, for fixed quasiparticle momentum, our angular analysis of the plasmonic response reveals pronounced anisotropy in the frequency spectrum when $\delta$ is finite and the band inversion terms vanish.
This highlights the role of topological and band structure anisotropies in shaping the collective mode behavior. And at the same time, introduces experimental signatures of the band topology and the relative importance of the different model parameters in the behavior of the frequency of the plasmonic excitations as a function of the momenta $q$.
For the bilayer semi-Dirac system, our analysis reveals the emergence of an additional plasmon mode. The relative phase between these two collective excitations can be tuned externally by rotating the upper layer. When the layers are aligned, the plasmon modes are in phase, resulting in a symmetric response. In contrast, a relative rotation of $\pi/2$ induces an antisymmetric mode structure. This tunable phase relation between collective excitations is particularly appealing from an experimental standpoint, as it may be realized via mechanical rotation or controlled stacking of layered van der Waals heterostructures. Such configurability suggests that bilayer semi-Dirac systems could serve as a platform for active plasmonic control, enabling functionalities like phase-tunable resonances, plasmonic switching, or interferometric detection in nanoscale devices. The ability to modulate plasmon modes through symmetry and topology thus opens pathways for designing novel optoelectronic or quantum sensing applications, where phase coherence and anisotropic dispersion are key operational elements.
\acknowledgments 
This work has been supported by Agencia Estatal de Investigación (Spain), Grant PID2022-136285NB-C31. The work of AL war supported by DFG via SFB1277.
\section{appendix}\label{appendix1}
We show the details for the calculation of the polarization function for a single-layer of semi-Dirac system described by the Hamiltonian in \eqref{eq1}. For this purpose, we begin by the expression for the overlapping factor, which is explicitly given as follows:
\begin{widetext}
\begin{equation}
F^{s s'}(\boldsymbol{k},\boldsymbol{k'})=|\langle \boldsymbol{k},s|\boldsymbol{k'} s' \rangle|^2=\frac{1}{4 E_k E_{k'}}\left|\sqrt{(E_k+s\Delta_k)( E_{k'}'+s'\Delta_{k'})}+ss'e^{i(\varphi_{k'}-\varphi_k)}\sqrt{(E_k-s\Delta_k)(E_{k'}-s'\Delta_{k'})}\right|^2,
\end{equation}    
\end{widetext}
As expected, in the limit $\Delta_k\rightarrow 0$, one recovers the overlapping factor of the standard semi-Dirac Hamiltonian. 
We can get some physical insight into the role of the anisotropy by evaluating the zero-temperature polarizability at low momenta compared to the Fermi momentum $q\ll k_F$, where we can approximate
\begin{equation}
\Theta(E_k-E_F)-\Theta(E_k'-E_F)\approx -{\bf q}\cdot\frac{\partial E_k}{\partial\bf k}\delta(E_k-E_F)\equiv -{\bf q}\cdot{\bf v}_k\delta(E_k-E_F),
\end{equation}
along with $F^{ss'}(\boldsymbol{k},\boldsymbol{k'})\approx F^{ss'}(\boldsymbol{k},\boldsymbol{k})= \delta_{ss'}$.
Thus, within this momenta regime, we can approximate the polarizability as
\begin{equation}
\Pi(\boldsymbol{q},\omega)\approx-\frac{1}{4 \pi^2} \displaystyle\int d^2k \frac{{\bf q}\cdot{\bf v}_k}{\hbar \omega-{\bf q}\cdot{\bf v}_k}\delta(E_k-E_F),
\end{equation}
where  the velocity vector is given as ${\bf v}_k=(v_x({\bf k}),v_y({\bf k}))$, and reads explicitly
\begin{equation}
{\bf v}_k=\left(\frac{2k_x(k_x^2+\delta-m_1\Delta_k)}{E_k},\frac{k_y(1-2m_1\Delta_k)}{E_k}\right)
\end{equation}
We can use the auxiliary vector ${\bf u} =E_k{\bf v}_k$ and the properties of the Dirac delta function to write alternatively
\begin{equation}
\Pi(\boldsymbol{q},\omega)\approx-\frac{1}{4 \pi^2} \displaystyle\int d^2k \frac{{\bf q}\cdot{\bf u}}{\hbar \omega E_F-{\bf q}\cdot{\bf u}}\delta(E_k-E_F),
\end{equation}
 To analytically solve the integral in $k_y$, we write
 \begin{equation}
\delta( E_k-E_F)=\delta(h(k_y^2))=\frac{\delta(k_y^2-k_+^2)}{|h'(k_+^2)| }+\frac{\delta(k_y^2-k_-^2)}{|h'(k_-^2)| }
 \end{equation}
where $h(k_\pm^2)=0$ and whenever $m_1\neq 0$, these roots read explicitly as{\footnote {The case $m_1=0$ is degenerate and leads to $k_\pm^2=E_F^2-m_0^2-(k_x^2+\delta)^2$}}:
\begin{equation}
k_\pm^2=\frac{-[1-2m_1(m_0-m_1k_x^2)]\pm g(k_x)}{2m_1^2},\quad m_1\neq 0
\end{equation}
 and, we have introduced the auxiliary function 
 \begin{equation}
 g(k_x)=\sqrt{1-2m_1(m_0-m_1k_x^2)-4m_1^2[(k_x^2+\delta)^2-E_F^2]}.
 \end{equation}
 Performing the explicit calculation, one gets 
 \begin{equation}
 |h'(k_\pm^2)|=\frac{g(k_x)|k_\pm|}{|E_F|}.
 \end{equation}
 Collecting these results, we can write then
 \begin{widetext}
  \begin{equation}
\delta( E_k-E_F)=\frac{|E_F|}{g(k_x)}\left(\frac{\delta(k_y-|k_+|)+\delta(k_y+|k_+|)}{2k_+^2}+\frac{\delta(k_y-|k_-|)+\delta(k_y+|k_-|)}{2k_-^2}\right)
 \end{equation}
 \end{widetext}
Using this expression, we can perform the $k_y$ integrals and get
\begin{widetext}
\begin{eqnarray}
\displaystyle\int_{-\infty}^{+\infty} dk_y \frac{{\bf q}\cdot{\bf u}}{\hbar \omega E_F-{\bf q}\cdot{\bf u}}\delta(E_k-E_F)&=&\frac{q_x|E_F|}{2g(k_x)}\left[\frac{u_{x+}}{k_+^2}\left(\frac{1}{\hbar\omega E_F-q_xu_{x+}-q_yu_{y+}}+\frac{1}{\hbar\omega E_F-q_xu_{x+}+q_yu_{y+}}\right)\right]+\\\nonumber
&&\frac{q_x|E_F|}{2g(k_x)}\left[\frac{u_{x-}}{k_-^2}\left(\frac{1}{\hbar\omega E_F-q_xu_{x-}-q_yu_{y-}}+\frac{1}{\hbar\omega E_F-q_xu_{x-}+q_yu_{y-}}\right)\right]+\\\nonumber
&&\frac{q_y|E_F|}{2g(k_x)}\left[\frac{u_{y+}}{k_+^2}\left(\frac{1}{\hbar\omega E_F-q_xu_{x+}-q_yu_{y+}}-\frac{1}{\hbar\omega E_F-q_xu_{x+}+q_yu_{y+}}\right)\right]+\\\nonumber
&&\frac{q_y|E_F|}{2g(k_x)}\left[\frac{u_{y-}}{k_-^2}\left(\frac{1}{\hbar\omega E_F-q_xu_{x-}-q_yu_{y-}}-\frac{1}{\hbar\omega E_F-q_xu_{x-}+q_yu_{y-}}\right)\right],
\end{eqnarray}    
\end{widetext}
where $u_{x\pm}=2k_x[(1+m_1^2)k_x^2+\delta-m_1(m_0-m_1k_\pm^2)]$ and $u_{y\pm}=k_\pm[1-2m_1(m_0-m_1k_x^2)+2m_1^2k_\pm^2]$, showing that $u_{x\pm}$ ($u_{y\pm}$) is an even (odd) function of $k_\pm$. However, $u_{x\pm}$ ($u_{y\pm}$) is an odd (even) function of $k_x$. Thus, one gets along the $q_y=0$ direction
\begin{widetext}
\begin{equation}
\Pi(q_x,\omega)\approx-\frac{q_x|E_F|}{4 \pi^2} \displaystyle\int_{-\infty}^{+\infty} dk_x\frac{1}{g(k_x)}\left[\frac{u_{x+}}{k_+^2}\left(\frac{1}{\hbar\omega E_F-q_xu_{x+}}\right)+\frac{u_{x-}}{k_-^2}\left(\frac{1}{\hbar\omega E_F-q_xu_{x-}}\right)\right].
\end{equation}    
\end{widetext}
One can perform a similar calculation along the $q_x=0$ direction, which leads to
\begin{widetext}
\begin{equation}
\Pi(q_y,\omega)\approx-\frac{q_y^2|E_F|}{4\pi^2} \displaystyle\int_{-\infty}^{+\infty} dk_x\frac{1}{g(k_x)}\left(\frac{u^2_{y+}}{k_+^2[(\hbar\omega E_F)^2-q_y^2u_{y+}^2]}+\frac{u^2_{y-}}{k_-^2[(\hbar\omega E_F)^2-q_y^2u_{y-}^2]}\right).
\end{equation}    
\end{widetext}

\end{document}